\documentclass[11pt]{article}

\usepackage{amsmath}
\usepackage{amssymb}
\usepackage{graphicx}
\usepackage{color} 
\usepackage[round,sort]{natbib}
\usepackage{soul}
\usepackage[labelfont=bf,labelsep=period,justification=raggedright]{caption}

\date{}

\definecolor{myYellow}{rgb}{1,0.95,0.75}
\newcommand{\hly}{\sethlcolor{white}\hl}

\newcommand{\mytilde}{\raise.17ex\hbox{$\scriptstyle\mathtt{\sim}$}}

\begin{document}
\bibliographystyle{plainnat}

%
%
%
%
%
%
%

\begin{flushleft}
{\Large
\textbf{Entrainment of the intrinsic dynamics of single isolated neurons by natural-like input}
}
\\
Asaf Gal$^{1,2,\ast}$, 
Shimon Marom$^{1}$
\\
\bf{1}  Network Biology Research Laboratories, Lorry Lokey Interdisciplinary Center for Life Sciences and Engineering, Technion, Haifa, Israel
\\
\bf{2} The Interdisciplinary Center for Neural Computation (ICNC), The Hebrew University, Jerusalem, Israel

$\ast$ E-mail: asaf.gal@mail.huji.ac.il
\end{flushleft}

\section*{Abstract}
Neuronal dynamics is intrinsically unstable, producing activity fluctuations that are essentially scale-free. Here we show that while these scale-free fluctuations are independent of temporal input statistics, they can be entrained by input variation.  Joint input output statistics and spike train reproducibility in synaptically isolated cortical neurons were measured in response to various input regimes over extended time scales  (many minutes).   Response entrainment was found to be maximal when the input itself possesses natural-like, scale-free statistics.  We conclude that preference for natural stimuli, often observed at the system level, exists already at the elementary, single neuron level.

\section*{Introduction}
Variability is a most prominent property of neural activity and neural response: neurons and neural networks behave in an irregular and indeterministic manner both spontaneously, and in response to series of stimuli. At the single neuron level, variability is observable in practically all aspects of evoked activity: irregularity of the spike train, trial-to-trial variability in spike counts, as well as  irreproducibility of train structure evoked by identical input series \citep{Faisal2008,Yarom2011}. Generally, variability in responses to repeated presentation of a stimulus is a significant constraint on information carrying and processing capacity.  However, as demonstrated in several cases, response variability might be quenched by a variation introduced to the input itself \citep{Churchland2010}.  

At the single neuron level, it was demonstrated that when stimulated with constant input, the neuronal spike train differ substantially from trial to trial \citep{Bryant1976,Mainen:1995qr}.  In contrast,  when stimulated with a fluctuating (filtered white noise) input, the reproducibility of the spike train is dramatically improved to the point of perfect repeatability, locking itself to (i.e.~entrained by) input fluctuations, reliably encoding its structure.   This key property was reproduced in a stochastic simulation of a Hodgkin-Huxley neuron, relating it to the properties of the underlying ion channels \citep{Schneidman:1998rc}. Other works have demonstrated the existence of repeatable spike patterns under different types of stimuli \citep{Fellous2004}. While these measurements and simulations were limited to timescale of seconds, it is known that when neuronal activity is observed over extended time scales, slower effects are recruited and excitability dynamics becomes rich \citep{Marom2010}.  In a recent work \citep{Gal2010a} we have shown that, indeed, when presented with long ($>1 h$) sequences of pulse stimuli, single neuron response dynamics becomes intermittent and irregular, exhibiting scale-free fluctuations (e.g. with auto-correlation that lacks a characteristic scale) and transitions between quasi-stable response pattern modes. Given these slower modulatory processes, it is not obvious that the statistically unstructured random input series, that are capable of quenching response variation over limited time scales, will effectively entrain response variability over extended durations (minutes and more).  The biophysical mechanism underlying the capacity of unstructured random input to entrain response variability relies on matching between time scales of input variations and time scales of the stochastic processes that generate the action potential.  In contrast, when longer stochastic processes are allowed, they are left unmatched by the above unstructured input.  It is therefore natural to hypothesize that in order to entrain neuronal response over extended time scales, the variations of the input series must match the scale-free temporal structure of intrinsic neuronal response dynamics. Indeed, at least at the system level, neuronal response variability is reduced under natural or natural-like sensory input \citep{Aertsen1981,Ruyter-van-Steveninck:1997gs,Baddeley1997a,Garcia-Lazaro2006,Yu2005,Garcia-Lazaro2011}. These natural-like signals are often characterized by long range temporal correlations, and a general scale-free temporal structure \citep{VOSS1975,DeCoensel2003,Simoncelli2003}. 

In this study we directly measure the impacts of input temporal structure on response variability over extended time scales in isolated cultured cortical neurons.  We show that while the response of neurons is temporally scale-free, independently of input statistics, entrainment is maximal when the input itself has a matching, scale-free structure. We also perform analogous analyses to those of Mainen and Sejnowski, quantifying the reproducibility of spike trains under different types of input.  Here too, natural-like input minimizes the trial-to-trial variability of the spike train. We conclude that the rich and complex neuronal dynamics enable the neuron to match its dynamics to that of the natural environment, and that ``tuning'' to natural input statistics arises already at the atomic level of neural processing. 

\section*{Materials and Methods}
\paragraph*{Culture preparation.}
Cortical tissues were obtained from newborn ($<24h$) rats (Sprague-Dawley) and dissociated following procedures described in earlier studies \citep{Marom:2002dg}. The cells were plated directly onto substrate-integrated multi-electrode arrays (MEA) and developed for a time period of 2-3 weeks before their use. A total amount of approximately $10^6$ cells was seeded on poly-ethyene-immine (PEI) coated MEAs. The preparations were kept in Minimal Essential Medium (MEM) supplemented with heat-inactivated horse serum (5\%), glutamine ($0.5mM$), glucose ($20mM$), and gentamycin ($10\mu g/ml$), and maintained in an atmosphere of 37$^{\circ}$C, 5\% CO$_{2}$ and 95\% air in an incubator as well as during the recording phases. An array of Ti/Au/TiN extracellular electrodes, 30$\mu$m in diameter, and spaced either 500$\mu$m or 200$\mu$m from each other (MultiChannelSystems - MCS, Reutlingen, Germany) were used. Synaptic transmission in the network was completely blocked by adding 20$\mu$M APV (amino-5-phosphonovaleric acid), 10$\mu$M CNQX (6-cyano-7-nitroquinoxaline-2,3-dione), and 5$\mu$M Bicuculline-methiodide to the bathing solution. \hly{All experiments were performed in accordance with the regulations (and under the supervision) of the Technion - Israel Institute of Technology animal care committee.} 
\paragraph*{Measurements and stimulation.}
A commercial amplifier (MEA-1060-inv-BC, MCS) with frequency limits of $150$-$3000Hz$ and a gain of x1024 was used.  Rectangular 200$\mu$s biphasic  $600$-$800mV$ voltage stimulation through extracellular electrodes was performed using a dedicated stimulus generator (STG4004, MCS). In the context of this study, no difference was observed in the behavior of neurons under current or voltage stimulation. Data was digitized to 16bit using a USB-ME256 system (MCS). Each recorded channel was sampled at a frequency of $20KHz$. One hour after the addition of synaptic blockers, the stimulation electrode was selected as one evoking well-isolated spikes with high signal to noise ratio, in as many recording electrodes as possible. From the selected recording electrodes, voltage traces of $15$-$20ms$ post stimulus were collected. Spike detection was performed off-line by a manual threshold-based procedure. A $3ms$ long spike shape was extracted for each response for further noise cleaning and analysis. Stability of spike shape and activity dynamics criteria were applied in order to validate experimental stability, as described in \citep{Gal2010a}. 
Random stimulation sequences were generated by modulating a constant stimulus interval sequence with a noise signal. This noise signal was either a white Gaussian noise or $1/f$ Gaussian noise generated by weighting the frequency components of the Gaussian white noise. In both cases a low cutoff was applied to have a minimum interval of $20ms$. \hly{The SD of the noisy stimulation interval sequence was set such that its CV will match the CV of its response constant interval. For example, if a neuron responded to a 100ms constant interval sequence with an average inter spike interval of 200ms and SD of 40ms, the SD of the noisy interval sequence was set to 20ms.} 

\paragraph*{Data analysis.}
Analysis throughout this study was performed on either the spike time series, or on a smoothened firing rate time series, produced by filtering the spike train with a sliding rectangular window (Figure~1C). Spectral analysis was performed on a binned time series, with a bin size of $1s$. The Power Spectrum Density (PSD) was estimated using a modified Periodogram. The Allan variance, which is commonly used to identify fractal point processes  \citep{Lowen:1996zv,Lowen:2005fx} was calculated by binning the time series with different bin sizes $T$. For each binned time series, the Allan variance is defined as $A(T)=\frac{<Z_T(k)-Z_T(k+1)>^2}{2\cdot <Z_T>}$, where $Z_T$ is the binned series. 
Another measure, widely used to characterize the temporal statistics of long memory and non stationary time series, is Detrended Fluctuation Analysis \citep[DFA, ][]{peng1995quantification}, in which the fluctuations (in terms of mean square error) around a piecewise linear fit to the time series are quantified, for different segment durations. 
Since both the Periodogram and the Allan variance are can be regarded as power law for the purpose of this work, power law functions were fitted to the tail of these curves, and their exponents were used as descriptive statistics \citep[see][]{Gal2010a}. 

The similarity between two spike trains was assessed with two distance measures: (1) A correlation based metric, defined as one minus the correlation between binned time series ($Z_a$,$Z_b$):  $d=1-corr(Z_a,Z_b)$.  This is a rate-based measure, which is insensitive to temporal features of the spike train below resolution dictated by the bin size. (2) The Victor-Purpura spike train metric
\citep{Victor2005,DVictor2009,Toups2011}, which defines the distance between the two spike trains as the minimal cost of transforming one into the other. Briefly, a spike train is modified by a combination three possible steps: inserting or deleting a spike, with a cost of +1, and moving a spike in time, with a cost of $q\cdot$$dt$ , where $q$ is a temporal resolution parameter. The value of $q$ sets the sensitivity of the metric to fine temporal features, and is set here to a default value of $q=1s$. The dependence of the results on the value of $q$ is shown in Figure 4. The Victor-Purpura metric was calculated using Matlab code downloaded from the website of Jonathan Victor.

\section*{Results}
\subsection*{Unexplained response variability is minimized by scale-free input}

In this paper we investigate the response properties of individual neurons, independent of synaptic and network effects.  To that end, experiments are conducted on cultured cortical neurons, functionally isolated from their network by means of pharmacological block of both glutamatergic and GABAergic synapses \citep[see][and Materials and Methods]{Gal2010a}.  Individual neurons are stimulated with sequences of short, identical extracellular electrical pulses. In response to a single pulse, a neuron either responds by emitting a spike, or fails to do so (Figure~1A); neuronal responses are monitored by extracellular recording electrodes. As previously reported \citep{Gal2010a}, when repeatedly stimulated over extended durations (minutes and more) with a low (\mytilde1Hz) stimulation rate, a neuron responds to each and every stimulation pulse (1:1 mode). As stimulation rate increases, the excitability of the neuron declines, and at a certain, neuron-specific critical stimulation rate, the 1:1 response mode breaks, and the neuron exhibits rich and complex response dynamics (Figure~1B).

Here, we study the properties of response dynamics evoked by three statistically different regimes of stimulation series:  (i) \textit{constant interval regime}, (ii) \textit{white noise regime} in which the interval series is modulated by a Gaussian white noise process, and (iii) \textit{scale-free regime} in which the intervals are modulated by a $1/f^\beta$ process, with parameter $\beta=1$, which is more representative of natural sensory input \citep{VOSS1975,DeCoensel2003,Simoncelli2003} and similar to activity properties of cortical neurons \textit{in-vivo} \citep{Lowen:1993yb,Lowen:1996zv,Teich:1997ta,Lowen:2001if,Bhattacharya2005}. The term \emph{scale-free} is used here to designate a timeseries with an autocorrelation that decays slowly, usually as a power-law without a typical scale. White noise on the other hand, which has a delta-function autocorrelation, is regarded as a zero scale signal. The choice of $\beta=1$ is typical to a wide range of natural signals, in natural environments and in biological systems. \hly{As will be shown, the results of this work are not sensitive to the \emph{exact} value of $\beta$, consistent with previous studies} \citep{Garcia-Lazaro2006}. Interval sequences in all stimulation regimes were normalized to have the same mean and standard deviation (the latter is applicable only to the second and third regime). The mean stimulation rate (the reciprocal of the mean stimulation interval) was set to a high enough value to drive the neuron beyond its critical point, leading to response failures and intermittency \citep{Gal2010a}. The interval standard deviation (SD) was chosen to approximately match the intrinsic SD of the response to constant interval stimulation.  Figure~1D shows examples of extracts from the three stimulation regimes, as well as their autocorrelation functions and the Power Spectral Densities (PSD).

Our analysis starts by stimulating neurons with long sequences  ($>1 h$) of each of the stimulation regimes. Figure~2A shows an example of the response of the same neuron to the three regimes.  It is immediately obvious that the three input regimes did not cause a significant difference in the statistical properties of the responses.  This is formally shown in the plots of the PSD, Allan variance, detrended Fluctuation Analysis (DFA), rate histograms and inter-spike interval (ISI) histograms (Figure~2B-F). Clearly, the macroscopic properties of the evoked spike trains are the same under all stimulation regimes:  the neuron exhibits scale-free dynamics, characterized by power law statistics, in accordance with previously published analysis \citep{Gal2010a}. 
In order to confirm the insignificance of the differences between the different statistical measures under the three regimes, a control experiment was performed, in which 4 neurons were stimulated and recorded, and each stimulation block was repeated 10 times. For each statistical analysis presented in Figure~2B-F, the relevant parameter was estimated independently from the response to each repetition. The value ranges of these parameters from a single neuron are depicted in the insets of each panel, and show that indeed the temporal statistics are the same under the three regimes.  We emphasize that the claim made here is not that the specific model chosen for each statistics is the correct one (i.e.~that the PSD is an exact power law), rather that these fits are good enough representative shapes, useful for comparing the population statistics.
The above results imply that the various membrane and cellular processes underlying the stochastic response fluctuations are effectively insensitive to the statistical structure of the input regime.  

$1/f$-type response statistics can be interpreted as a modulation of neuronal excitability by a cascade of oscillating processes at various time scales. An oscillator can be entrained (i.e. phase-locked) by a driving stimulus at a frequency that matches the oscillator's natural frequency, and with a magnitude proportional to the oscillation amplitude. This suggests that the $1/f$ intrinsic fluctuations could be entrainable by a matching $1/f$ stimulation.

Figure~3 demonstrates this effect: The stimulation and response of a neuron are plotted in two timescales, for the three stimulation regimes. On the short timescale (panel A, $5s$ bin size), the response in the white noise and scale-free regimes nicely follows the stimulation, while in the constant interval regime, \hly{the input has no bearing on timing of the output fluctuations}. 
On a longer timescale (panel B, bin size of $100s$), the white noise input is practically flattened, becoming constant; as a result, it fails to entrain the response. In contrast, in the scale-free regime entrainment is evident throughout.
Figure~3C shows a scatter plot of the input and output rates of the neuron, calculated with a $5s$ bin size.  As expected from the data of Figure~2, the input and output ranges of the white-noise (red) and scale-free (green) are practically identical.  However, the correlation in the scale-free data is much higher, as expressed in the reduced variability around the linear trend line. This indicates that indeed the scale-free fluctuations in neuronal dynamics are best entrained by scale-free input. For 13 of the 17 recorded neurons, the input output correlation coefficient significantly increases under scale-free input regime, compared to white noise regime. \hly{A Wilcoxon signed-rank test yields $p<0.01$ for the effect of the input regime on input/output correlation.} Thus, under scale-free input, there is no change in the amount of variability in neuronal response, but more of it is explained by the input. 
Figure~3D shows a scaled correlation analysis: The correlation between input and output is calculated for responses in white noise and scale-free regimes on different timescales. The correlation in a given timescale $T$ is calculated by smoothing the stimulation and response series with a rectangular window of width $T$, and subtracting from it a version of the series, smoothed with a window of size $25T$. This effectively band-passes the time series around $T$. \hly{While the correlation in the short timescale regime is similar for the white noise and scale free regimes (the apparent increase for white noise is non typical, a Wilcoxon signed-rank test yields insignificant difference), for longer timescales it decreases in the case of white noise compared to scale free ($p<0.01$, Wilcoxon signed-rank test).}

\hly{The above results show that there is a significant increase in correlation upon change from white noise (i.e. $\beta=0$) to scale free ($\beta=1$) input. However, there is nothing unique about $\beta=1$: other types of inputs might serve just as well. In order to asses the sensitivity of the increased input-output correlation to the value of $\beta$, neurons (n=31) were exposed to 9 stimulation blocks, each characterized by a different value of $\beta$, ranging from 0 to 2, and lasting 70 minutes. For each block, the correlation between input and response was computed, and compared with the correlation for $\beta=0$. The population statistics of this correlation ratio are depicted in} Figure~3E, \hly{and show a clear concave shape, with a peak in the mid-range values, and a decrease toward lower and higher values of $\beta$. A similar peak is observed when plotting the distribution of preferred exponents (i.e. the exponents which results in the strongest correlation for each neuron,} Figure~3F).  \hly{These peaks around $\beta=1$, however, are wide, suggesting that the variability between neurons is considerable, and that inputs characterized by a wide range of exponents are equally effective in entraining neuronal responses.}

\hly{The decrease of the correlation for input with large exponents, which are dominated by slow oscillations, is important for the understanding of this phenomena.} Figure~3G \hly{shows extracts from the response to $1/f$ stimulation (left) and $1/f^2$ stimulation (right). It is easy to see how, for $1/f^2$, the faster fluctuations remain un-entrained, while slower oscillations are nicely locked by modulations in the input.}

\subsection*{Neuronal response repeatability is maximized by scale-free input}

Another functionally relevant aspect of the entrainment described above, concerns response repeatability.  It is well established that in spite of the extensive variability of neuronal responses,  spike train structure is repeatable when the input is fluctuating, both at the single neuron and sensory system levels \citep{Mainen:1995qr,Ruyter-van-Steveninck:1997gs,Churchland2010}. In the following set of experiments we ask whether a scale-free stimulation regime enhances repeatability in general, and over long time scales in particular.  To this aim we stimulated neurons with 10 repetitions of the same input sequence under each of the three stimulation regimes: constant intervals, white noise and scale-free.  Each sequence lasted 10 minutes, separated by a 10 minute break.  Figure~4A shows responses of one neuron to ten identical sequences under each stimulation regime. It is immediately obvious that although some reproducibility exists under the white noise regime, the reproducibility of responses to scale-free sequences is much higher. This is a direct consequence of the previous section analysis: if responses are more correlated to the input, they will be more correlated between themselves under repetitions of the same input sequence.  This observation is quantified using two kinds of spike train similarity measure: a rate-based measure (the correlation distance between spike histograms, calculated with 1s bins; see Methods section), and a time-based measure  \citep[the Victor-Purpura distance, ][]{Victor2005,DVictor2009,Toups2011}. Figure~4B depicts the pairwise distance matrices of the responses of Figure~4A, calculated using these two metrics. Responses to white noise stimulation are significantly more reproducible than responses to constant input; this is in agreement with \citep{Mainen:1995qr}.  But as expected from the results presented above (Figure 3) the responses to the scale-free sequences are significantly more reproducible than those of white noise input.  The purple lines of Figures~4C and 4D depict the mean and SD of the values in the distance matrices shown in Figure 4B.  Of 14 recorded neurons, 11 neurons showed significant improvement in reproducibility for scale-free sequences compared to constant interval and white noise, as quantified by the two metrics (there were no cases of disagreement). The grey lines depict the trends of mean distances calculated for all of the 14 recorded neurons.  \hly{A Wilcoxon signed-rank test for the mean pair-wise distances shows an overall significant effect for the input regime on response repeatability ($p<0.01$ for both white noise vs. constant and for scale-free vs. white noise, for both metrics).}

The results are insensitive to the choice of the temporal parameter $q$ of the spike train metric (Figure~4F), which may point to the lack of characteristic scale in this phenomenon.

The above enhanced reproducibility does not stem from a decreasing spike rate. Next to the preservation of the mean stimulation rate in all sequences, the spike rate itself does not decrease for scale-free input (see Figure~4E). For all of the 14 cells recorded, there was no significant decrease in firing rate under the scale-free regime compared with constant and white noise.

Since the scale-free input itself is structured, it is expected that even a ``Bernoulli'' neuron that has a constant probability of response to a pulse stimulation, will have some reproducibility of its output spike train. Figure~5A shows a comparison between metric analyses on actual responses (grey), and on a surrogate Bernoulli neuron (purple) that responds with a constant probability (set to the mean response probability of the real neuron).  As expected, the responses of the simulated neuron are indeed more reproducible for scale-free input, but not as consistent as the real neuron. 
It is also possible to construct a more detailed neuronal response curve, which takes into account the dependency of the response probability on the last interval between stimuli (an example for such a curve is given in Figure~5B). Interestingly, the curves of the white noise and scale-free regimes substantially differ, pointing to a strong history dependence in response probability. As shown in Figure~5A, the resulting metric analyses (orange) behave more like the real neuron when compared to the Bernoulli neuron, yet cannot account for the entire effect. It is reasonable to believe that one might construct a response model that takes into account deeper history of the stimulation and response sequences, to produce better fitting.  It should be emphasized though, that while these neuronal response models do reproduce the repeatability effect to a significant extent, they do not reproduce the intrinsic scale-free fluctuations, and can not be considered as successful explanatory models.

A scale-free input is characterized by the abundance of relatively long ``breaks'' in stimulation, or long periods with low stimulation rate, enabling recovery of internal processes from previous activations.  As repeatedly shown over the past 15 years, the longer a neuron is exposed to repeated activations, longer recovery times are required \citep{Toib:1998uk,Ellerkmann2001,Fairhall:2001tw,Lundstrom:2008kb,Marom:2009ov}.  A scale-free input statistics is inherently matched to such a mechanistic context:   elongating a scale-free input series naturally gives rise to longer breaks, hence allowing for stabilization on every scale.  An illustration of this property is provided by reanalyzing the data of Figure~4 over blocks of increasing lengths.  The results are summarized in Figures~5C and 5D, \hly{showing that the accumulation of variability (or divergence of response) with increasing block size (1-10 minutes range) is significantly slower than its accumulation in responses to other stimulation regimes.}

\section*{Discussion}
In this paper we have shown how a natural-like, scale-free input entrains fluctuations of single neuron responses over extended timescales. We have demonstrated this property by comparing neuronal responses in three different stimulation regimes: \textit{constant interval}, \textit{white noise} and \textit{scale-free}. In the case of the scale-free regime, the correlation between the input and the response is significantly higher, and the repeatability of response is considerably enhanced. These characteristics are stable over long, practically unlimited durations. \hly{While the results do show a preference to mid-range values of $\beta$ (around 1), there is nothing special about the exact value; what seems to be important is that the entrainment decreases when the slow frequency component in the input becomes either too dominant (large $\beta$) or marginal (low $\beta$).}

It has long been acknowledged that responsiveness of neural systems is optimized to the ranges of statistics found in natural inputs \citep{Aertsen1981,Ruyter-van-Steveninck:1997gs,Baddeley1997a,Garcia-Lazaro2006,Yu2005,Garcia-Lazaro2011}.  Here we show that preference to natural statistics is not limited to large-scale neural systems;  rather, it goes all the way down, to the atomic level of neural organization, namely the single, isolated neuron.

At the shorter timescales, the entrainment of neuronal fluctuations by white noise input is explained by the fast stochastic processes underlying the generation of action potentials \citep{Schneidman:1998rc}. There, a variation in input allows for recovery of inactivation processes, unlike the case of a constant input that drives the neuron to operate around the limit of channel availability threshold, making it highly sensitive to stochastic events.  It is reasonable to assume that a similar explanation would also be appropriate over extended timescales:  a long break (or a period of low-rate stimulation) enables recovery of slow processes, in contrast to constant input (or shortly correlated input) that drives these processes to a highly stochastic operation point.  It is natural to assume that such processes include slow inactivation properties of the ionic channels themselves \citep{Toib:1998uk,Ellerkmann2001,Marom:2009ov,Soudry2010}, or other cellular modulatory processes (e.g.,~protein phosphorylation, protein synthesis and metabolic cycles). 

From the more abstract, functional point of view, when the temporal structure of the input is relatively dull, it can only entrain a narrow range of cellular processes underlying neuronal dynamics.  Under these conditions, a large fraction of the response variability is tagged ``unexplained''.  However, when the input is temporally rich, it matches the temporal manifold complex structure of the intrinsic dynamics, and the former ``unexplained'' variability becomes information-carrying.    
Thus, a neuron can be viewed as a collection of entangled information channels, distributed over a continuum of timescales.  In this picture, information transfer is maximized when there is information to be transferred on any given scale.

\subsection*{Acknowledgments}
The authors thank  Danni Dagan, Danny Eytan and Avner Wallach for helpful discussions and insightful input, and Eleonora Lyakhov and Vladimir Lyakhov for invaluable technical assistance in conducting the experiments. The research leading to these results has received funding from the European Unions Seventh Framework Program FP7 under grant agreement 269459 and was also supported by a grant of the Ministry of Science and Technology of the State of Israel and MATERA grant agreement 3-7878.

\newpage

\section*{Figure captions}

\textbf{Figure 1.} Data analysis and stimulation regimes. \textbf{(A)} Examples of voltage traces recorded following several stimulation pulses, delivered at $15Hz$. In response to such a pulse, a neuron sometimes emits a spike (blue traces) or fail to do so (red traces). Stimulation pulses are 400$\mu$s wide and start at $t=0$. \textbf{(B)} Color representation of response traces. Each line represents a single response trace. Responses to consecutive stimulation pulses, delivered at $15Hz$, are ordered top to bottom. Voltage is color coded, red for high voltage and blue for low. It can be seen that spikes are fired in response to some of the stimulations, in a seemingly random and complex manner. 
\textbf{(C)} Examples for stimulation sequences, each of $1h$ length. The signals shown are stimulation rates, the reciprocals of the stimulation interval series.  Examples are given of the three stimulation regimes (see Results section): constant interval (upper panel, blue), white noise (middle panel, red) and scale-free (bottom panel, green). 
\textbf{(D)} The autocorrelation function for the white noise (red) and scale free (green) sequences.
\textbf{(E)} Power Spectral Density (PSD) for the white noise (red) and scale free (green) sequences.

\vspace{1cm}

\textbf{Figure 2.} Response statistics under the three stimulation regimes. Main panels show raw data and statistical measures of a typical neuron to the three regimes ($1h$ of stimulation for each sequence): constant interval (blue), white noise (red) and scale-free (green). Insets show results from a neuron subjected to a control experiment where the significance of the differences in responses to each of the regimes are assessed by 10 repetitions of each stimulation block.
\textbf{(A)} Extracts from the firing rate of the neuron in response to the three stimulation types ($8min$ length, $1s$ bin width). 
\textbf{(B)} Histograms of firing rate values from one repetition, calculated with $5s$ bin size. The inset shows the distribution of  mean firing rate value from each of the 10 repetitions in the control experiment. Filled box represents the 25\%-75\% range, and whiskers extend to the extreme values.  
\textbf{(C)} Histograms of ISI values from one repetition. The inset plots the range of values of the ISI coefficient of variation under 10 repetitions.  
\textbf{(D)} Response PSD from one repetition, on double logarithmic axes. The inset depicts the range of exponent values for a power law fit to the low frequency tail of the PSD.   
\textbf{(E)} Detrended Fluctuation Analysis (DFA, see Methods) of the responses from one repetition. The inset depicts the range of exponent values for a power law fit to the Fluctuation curve. 
\textbf{(F)} Allan variance of the responses  of the responses from one repetition. The inset depicts the range of exponent values for a power law fit to the linear tail of the curves. 

\vspace{1cm}

\textbf{Figure 3.} Input and output correlations under the three stimulation regimes.
\textbf{(A)}  Extracts from the response rates of a neuron to long stimulation (1 h) under the three regimes. Responses and stimulation are binned with a 5s bin size, 30 bins are shown. Mean is subtracted to have the input and output aligned. The responses under white noise and scale-free regimes follow closely the stimulation (black line), while for constant interval the variability is freely running. 
\textbf{(B)} Extracts from the same experiment, using a 100s bin size, 30 bins are shown. The response in the scale-free regime is still locked to stimulus fluctuations, while input fluctuations in the white noise regime are substantially diminished, therefore unable to lock response variability. 
\textbf{(C)} A scatter plot of the stimulation rate against response rates, \hly{calculated with a 5s bin size}. White noise input in red, scale-free input in green. Mean is removed for visual clarity, and both axes are in standardized units. While the marginal distributions for both inputs are similar, the correlation in the scale-free case is significantly higher.
\textbf{(D)} Scaled correlation analysis. The correlation between input and output rates are calculated in different timescales, for white noise and scale-free regimes. The correlation in a given timescale $T$ is calculated by smoothing the series with a rectangular window of width $T$, and subtracting a version of the series smoothed with a window of size $25T$, effectively band-passing the time series.
\hly{\textbf{(E)} Dependence of the input output correlation on the exponent $\beta$ of the input. Neurons (n=31) were stimulated with blocks of 70 minutes duration with exponent ranging from 0 to 2. The input output correlation ($c_\beta$) was calculated from the responses of each block, using a 1s time bin. The graph shows the population statistics of the ratio $c_\beta/c_0$ for each $\beta$, using box and whisker plot. The red horizontal line marks the median, the box marks the lower and upper quartiles and the whiskers the range of data values. Outliers (values outside the range of 3 SD units from the average value) are marked with red points. 
\textbf{(F)} Distribution of preferred exponent values (the exponents which results in the strongest correlation for each neuron). 
\textbf{(G)} Extracts comparing the entrainment for $\beta=1$ (left) and $\beta=2$ (right), from a single neuron, demonstrating lack of entrainment for the fast fluctuations for the latter case. Traces are smoothed with a 2s rectangular window, time scale is identical for the two extracts.}

\vspace{1cm}

\textbf{Figure 4.} Repeatability of neuronal response. \textbf{(A)} The firing rates of a neuron under 10 identical stimulation sequences, under the three regimes. Responses to the constant interval stimulation show no reproducibility at all. Responses to the white noise input show reproducibility on short timescales, as can be seen in the inset. At longer timescales there is again no response repeatability. The responses to the scale-free sequence are the most reproducible, and lock to the input on many timescales. Stimulation rates are plotted in black for the three plots, but are normalized to have the mean and SD of the responses for clear visualization. \textbf{(B)} Pairwise distance matrices between responses to repetitions of the same stimulation sequence, calculated according to correlation metric (left) and the Victor-Purpura (VP) distance with temporal parameter $q=1$ (right, see Methods section for details). Distance values are color coded, and for each metric the color scale was normalized to the maximal value. Diagonal pixels were whitened for visual clarity. \textbf{(C)} The purple line designates the mean and SD of the pairwise distances depicted in \textbf{B}, for the correlation metric, quantifying the change in repeatability under the three regimes.  Also shown are the means for all 14 recorded neurons (grey for those which showed significant improvement for scale-free input regime, brown for those which didn't). \textbf{(D)} Same as \textbf{C}, but for VP distance. \textbf{(E)} Mean and SD of the average firing rate per trial for the neuron of \textbf{A} (purple). The firing rate significantly increases for scale-free input for this neuron. While this is not generally true across the population of neurons recorded, the firing rate never significantly decreases for scale-free input. \textbf{(F)} The effect of the temporal scale $q$ on the results for the spike train metric. $q$ is varied from $0.25s$ (light grey) to $10s$ (dark grey). While the typical values differ, the effect exists for any choice of $q$, as can be expected from the lack of typical scale for the phenomena. The red curve is for $q=1$.

\vspace{1cm}

\textbf{Figure 5.} \textbf{(A)} Metric analysis on surrogate data. As explained in the Results section, the purple curve is metric analysis results, as in Figure 4, on surrogate data generated by applying a Bernoulli response model with a mean equaling the average response of the neuron, using VP distance with $q=1$. The orange curve is results of the same analysis applied to data generated by a Bernoulli response model conditioned on the last interval, according to the curve in \textbf{B}. The grey curve is the actual experimental results for this neuron. \textbf{(B)} The response probability of a neuron, conditioned on the last stimulation interval, for white noise input (red, lower curve) and scale-free input (green, upper curve). Both curves, as expected, are mostly increasing. The difference between the curves implies the dependence of the probability on history longer than the last interval. Since the scale-free input contains correlations between intervals, its response curve differs from the white noise curve.
\textbf{(C)} The effect of block length. The metric analysis of Figure~4 was repeated with various block sizes, i.e. with analysis performed on the first $T$ seconds of each repetition, using the correlation metric. Data was taken from the onset of the stimulation block, including the transient phase. This analysis shows how the responses to the constant interval input and white noise input are drifting apart relatively rapidly (mean distance increases), while the responses to the scale-free input are forced together by the input dynamics, and show only a slow and moderate gain in distance. 
\textbf{(D)} Block length analysis as in C, using VP distance . 
It should be pointed that both the metrics used here are effectively normalized to the input length, in contrast to metrics like Euclidean and others which are extensive in input length.  

\clearpage

\begin{figure}
\centerline{\includegraphics[width=8.5cm]{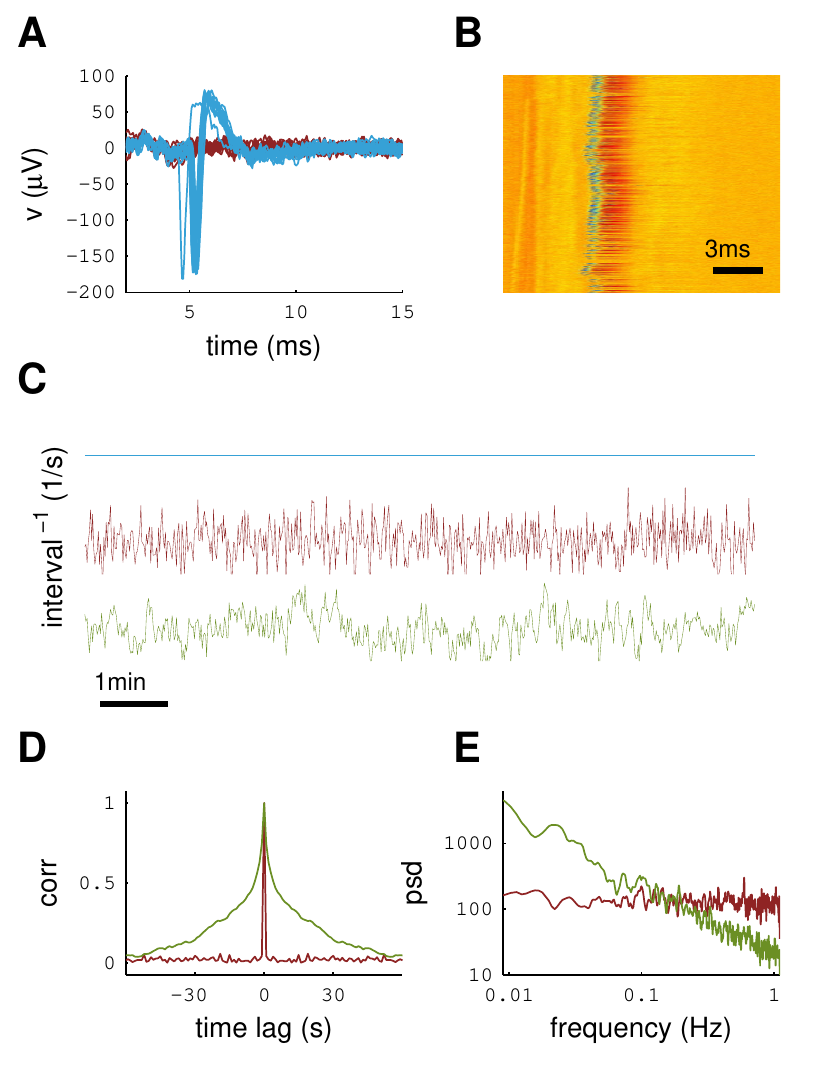}}
\caption{}
\label{fig12Methods}
\end{figure}

\clearpage

\begin{figure}
\centerline{\includegraphics[width=8.5cm]{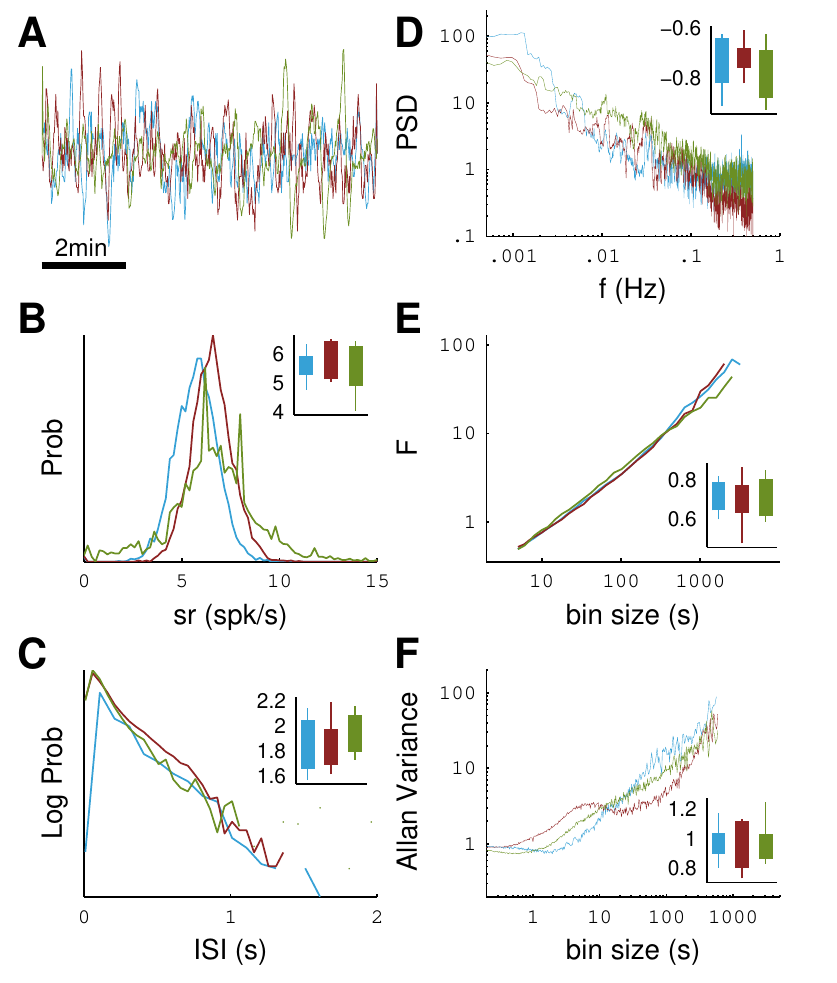}}
\caption{}
\label{figMarginalStatistics}
\end{figure}

\clearpage

\begin{figure*}
\centerline{\includegraphics[width=8.5cm]{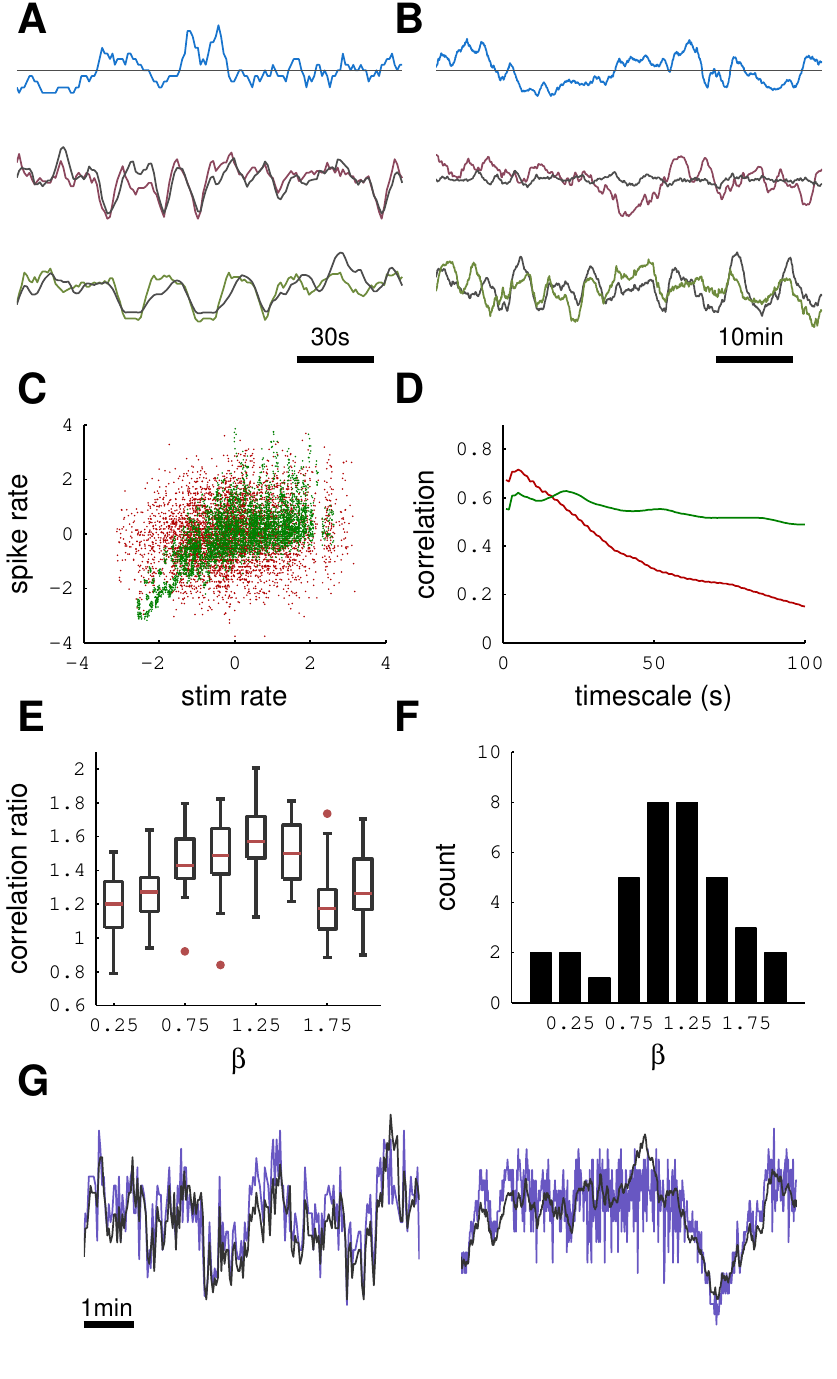}}
\caption{}
\label{figJointStatistics}
\end{figure*}

\clearpage

\begin{figure*}
\centerline{\includegraphics[width=17.6cm]{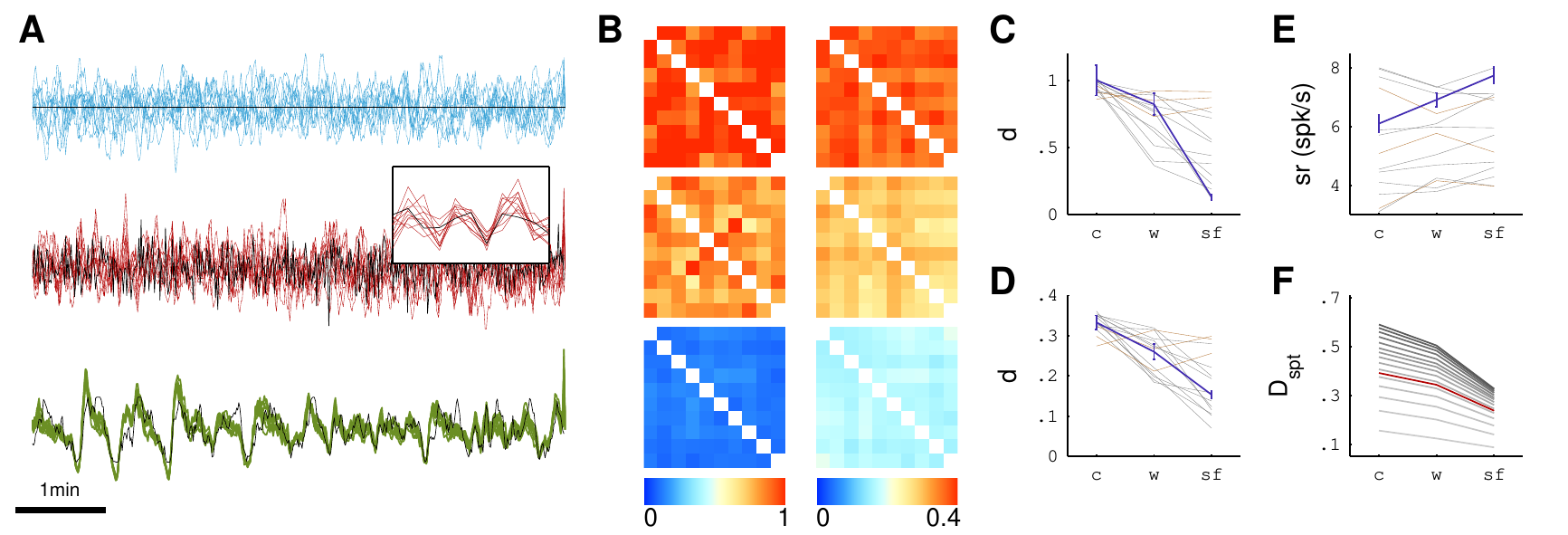}}
\caption{}
\label{figRepeatability}
\end{figure*}

\clearpage

\begin{figure}
\centerline{\includegraphics[width=8.5cm]{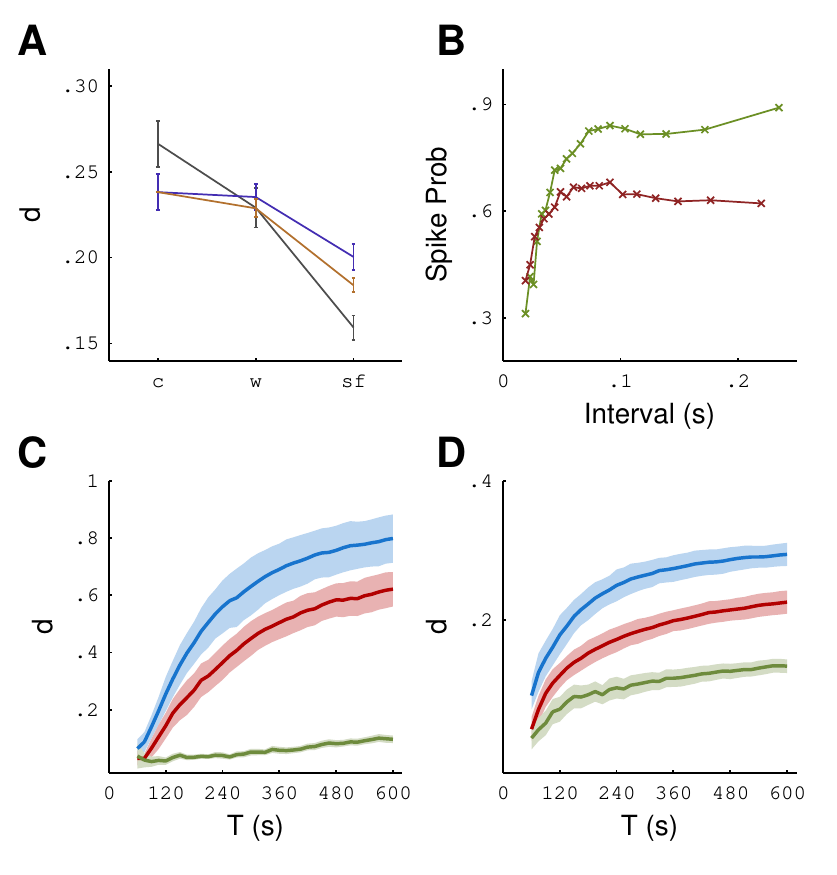}}
\caption{}
\label{fig6}
\end{figure}

\end{document}